\documentclass[amsfonts, amssymb, amsmath, preprint, aps, showkeys, nofootinbib, onecolumn, superscriptaddress]{revtex4-1}
\usepackage[hyperindex,breaklinks]{hyperref}
\usepackage{float}
\usepackage{calc}
\usepackage{dcolumn}% Align table columns on decimal point
\usepackage{bm}% bold math
\usepackage[T1]{fontenc}
\usepackage{braket}
\usepackage{amsmath}
\usepackage{amssymb}
\usepackage{graphicx}
\usepackage{xcolor}
\usepackage{lipsum}
\usepackage{nccmath}

\usepackage{setspace}
\usepackage[font = small,justification=justified, format=plain]{caption}
\graphicspath{ {./images/} }

\begin{document}
    \title{Collisional cooling of ultracold molecules}
    \author{Hyungmok Son}
        \email[Correspondence email address: ]{hson@g.harvard.edu}
        \affiliation{Research Laboratory of Electronics, MIT-Harvard Center for Ultracold Atoms, Department of Physics, Massachusetts Institute of Technology, Cambridge, Massachusetts 02139, USA}
        \affiliation{Department of Physics, Harvard University, Cambridge, Massachusetts 02138, USA}
    
    \author{Juliana J. Park}
        \affiliation{Research Laboratory of Electronics, MIT-Harvard Center for Ultracold Atoms, Department of Physics, Massachusetts Institute of Technology, Cambridge, Massachusetts 02139, USA}
        
    \author{Wolfgang Ketterle}
        \affiliation{Research Laboratory of Electronics, MIT-Harvard Center for Ultracold Atoms, Department of Physics, Massachusetts Institute of Technology, Cambridge, Massachusetts 02139, USA}
        
    \author{Alan O. Jamison}
        \email[Present address: ]{Department of Physics and Astronomy, Institute for Quantum Computing, University of Waterloo, Waterloo, ON, Canada}
        \affiliation{Research Laboratory of Electronics, MIT-Harvard Center for Ultracold Atoms, Department of Physics, Massachusetts Institute of Technology, Cambridge, Massachusetts 02139, USA}

    % \date{\today}
    
    \begin{abstract}
        Since the original work on Bose–Einstein condensation\cite{JILABEC,MITBEC}, the use of quantum degenerate gases of atoms has enabled the quantum emulation of important systems in condensed matter and nuclear physics, as well as the study of many-body states that have no analogue in other fields of physics\cite{ManyBodyRMP}. Ultracold molecules in the micro- and nanokelvin regimes are expected to bring powerful capabilities to quantum emulation\cite{DipolarQSimReview} and quantum computing\cite{DeMilleQComp}, owing to their rich internal degrees of freedom compared to atoms, and to facilitate precision measurement and the study of quantum chemistry\cite{ColdMolReview}. Quantum gases of  ultracold atoms can be created using collision-based cooling schemes such as evaporative cooling, but thermalization and collisional cooling have not yet been realized for ultracold molecules. Other techniques, such as the use of supersonic jets and cryogenic buffer gases, have reached temperatures limited to above 10 millikelvin[78]. Here we show cooling of NaLi molecules to micro- and nanokelvin temperatures through collisions with ultracold Na atoms, with both molecules and atoms prepared in their stretched hyperfine spin states. We find a lower bound on the ratio of elastic to inelastic molecule–atom collisions that is greater than 50---large enough to support sustained collisional cooling. By employing two stages of evaporation, we increase the phase-space density of the molecules by a factor of 20, achieving temperatures as low as 220 nanokelvin. The favourable collisional properties of the Na–NaLi system could enable the creation of deeply quantum degenerate dipolar molecules and raises the possibility of using stretched spin states in the cooling of other molecules.
    \end{abstract}

    \maketitle
	The full potential of ultracold atoms was not realized until the advent of collision-based cooling methods such as evaporative and sympathetic cooling. Although atomic systems have been recently used to demonstrate laser cooling to quantum degeneracy, these schemes still require collisional thermalization\cite{SchreckLaserCool,VuleticLaserCooling}. Therefore, there has been much work over the past 15 years\cite{HutsonFirstSymp} to achieve collisional cooling of ultracold molecules. Buffer gas cooling\cite{BufferGasCaH} cannot push below $100\;{\rm mK}$  owing to the rapidly diminishing vapour pressure of buffer gases at such temperatures. Supersonic expansion\cite{Supersonic_CO} can produce temperatures around $100\;{\rm mK}$. 
	Controlled collisions in crossed molecular beams\cite{Ar-NO_Collisions} can decrease the laboratory-frame velocity of particles while narrowing the velocity distribution. However, this technique has not been demonstrated below about $500\;{\rm mK}$. Merged supersonic beams can be used to study collisions at energies equivalent to a temperature of $10\;{\rm mK}$ (ref. \cite{mergedbeams}).
	
	Cooling below $100\;{\rm mK}$ calls for trapping molecules in magnetic or electrostatic traps and for good collisional properties (that is, a ratio of elastic to inelastic collisions much greater than 1). Such traps require preparing molecules in weak-field seeking states, which are never the absolute ground state, allowing inelastic state-changing collisions to eject the cold molecules from the trap. A variety of systems have been proposed for evaporative or sympathetic cooling of molecules \cite{HutsonFirstSymp,HutsonPolandSymp,HutsonHSymp,TarbuttCaFSymp,TscherbulSrFSymp,BohnCaOH}. So far, elastic collisions have been observed clearly in ${\rm O}_2$ at temperatures below $1\;{\rm K}$\cite{NareviciusO_2} and possibly in OH radicals around $10\;{\rm mK}$ (ref. \cite{JunOHRevised} corrects an earlier report \cite{JunOHEvap}). In the ${\rm O}_2$ case, inelastic collisions prevent thermalization and collisional cooling.
	
	In recent years, the assembly of molecules from ultracold atoms\cite{JinYeKRb} and the direct laser cooling of molecules\cite{DeMilleMOT, DoyleMOT, TarbuttMOT, JunMOT} have both expanded to new molecules and temperature regimes. These techniques have achieved molecular systems at temperatures less than $100\;{\rm mK}$, raising the challenge of collisional cooling in the micro- and nanokelvin regimes. Optical traps enable trapping of the absolute ground state, which removes the concern of state-changing collisions. However, collisional cooling in the absolute ground state of chemically stable systems has not yet been realized\cite{HutsonAlkaliChem}. By contrast, chemically stable molecular species have shown anomalously high inelastic loss rates that preclude collisional cooling, possibly owing to collision complex formation\cite{StickyBohn} or interactions with optical trapping beams\cite{KarmanMolLoss}.
	
	In this study we observe sympathetic cooling of rovibrational ground-state triplet $^{23}{\rm Na}^6{\rm Li}$ molecules by $^{23}{\rm Na}$ atoms, both of which are prepared in their upper stretched hyperfine spin states (that is, states with both nuclear and electronic spins aligned along the direction of the magnetic bias field). Although sympathetic cooling of one atomic species by another has been observed in various ultracold atomic mixtures\cite{WiemanSympBECs}, triplet NaLi was considered unlikely to have sufficiently good collisional properties to support such cooling. NaLi has energetically allowed chemical reactions, even in the electronic ground state (that is, a singlet state), and the triplet state has an electronic excitation energy of $0.8\;{\rm eV}$ or $10,000\;{\rm K}$.
    
    Furthermore, theoretical studies on various systems have explored the possibility of suppressing inelastic collisions and reactions by spin polarization, giving pessimistic predictions for triplet molecules and more favourable ones for doublet molecules\cite{NHChemReacts, collisionTheoryKrems, HutsonNHpN, TscherbulSrFSymp, TscherbulCaHLi}. Nonetheless, here we report clear thermalization and sympathetic cooling of triplet NaLi to 220 nK. We observe 20-fold increases of phase-space density (PSD), opening the possibility of collisional cooling to deep quantum degeneracy.
	\begin{figure}[H]
		\centering
		\includegraphics[width = 107mm, keepaspectratio]{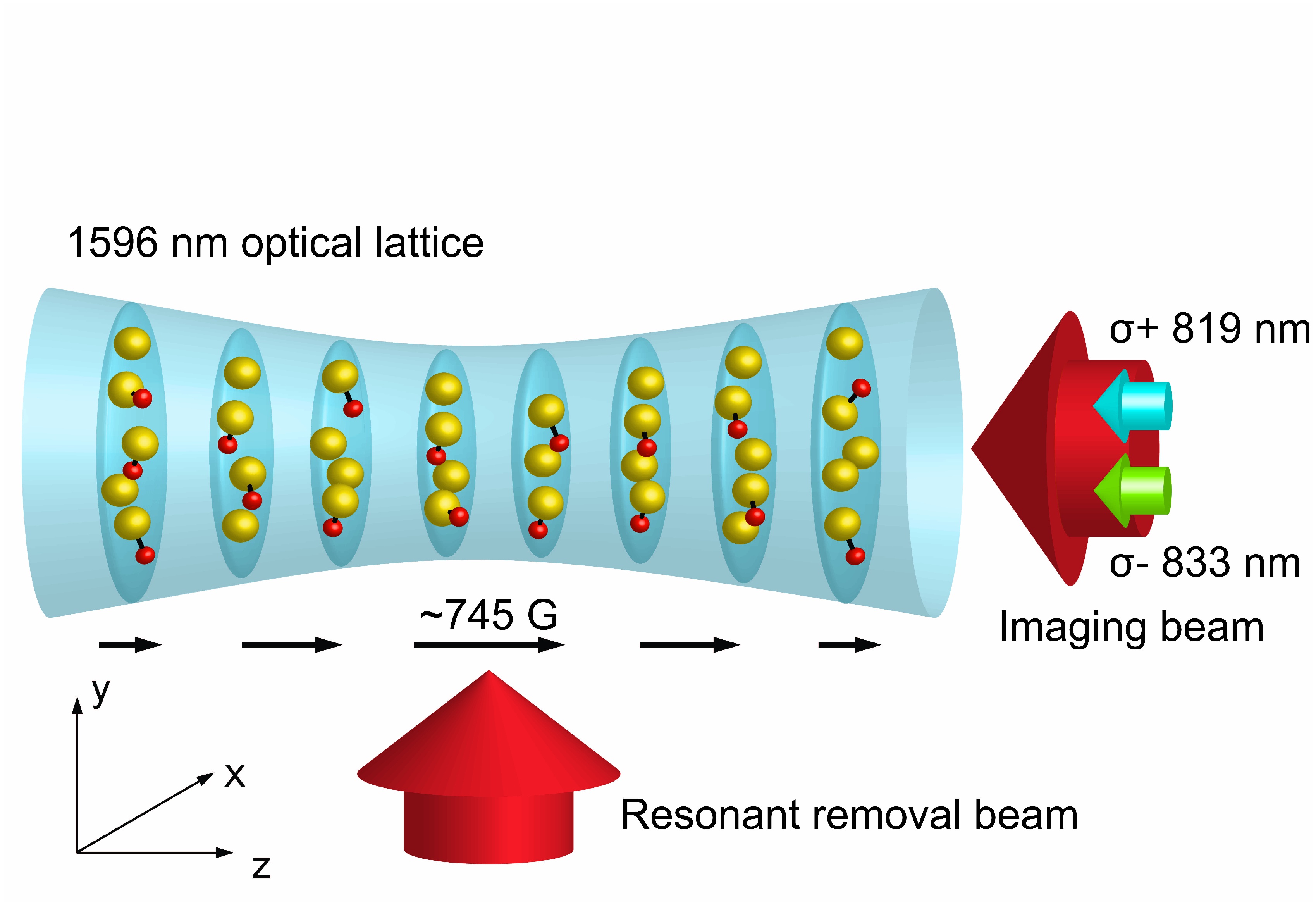}
		\caption{\textbf{Experimental setup.} The Na atoms (yellow circles) and NaLi molecules (yellow and red circles on black sticks) are trapped in a 1D optical lattice formed by a 1,596-nm laser, which is retro-reflected. The magnetic field, which defines the quantization axis, is coaxial with the lattice beam. The free atoms that remain after the formation of the ground-state molecules are removed by resonant light in the radial direction ($y$ axis in the figure). Stimulated Raman adiabatic passage (STIRAP) is performed with circularly polarized beams of two wavelengths (833 nm and 819 nm) that propagate along the axial direction ($z$ axis in the figure). $\sigma^{+}$ and $\sigma^{-}$ represent left-handed and right-handed circular polarization, respectively. The ground-state molecules are detected by absorption imaging along the axial direction of the atoms resulting from the dissociation of NaLi.}
		\label{fig:setup}
	\end{figure}
	\begin{figure}[H]
		\centering
		\includegraphics[width = 98mm, keepaspectratio]{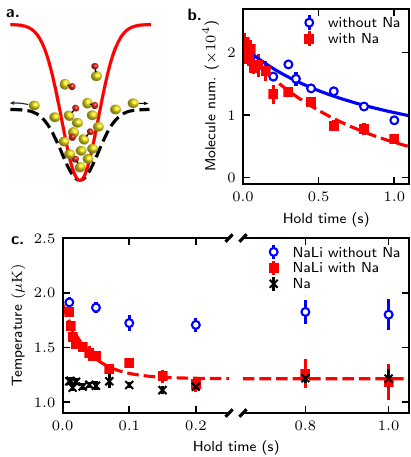}
		\caption{\textbf{Thermalization of Na and NaLi.} \textbf{a}, The trapping potential of molecules (red solid line) is deeper than that of Na atoms (black dashed line). This allows us to evaporate Na atoms with negligible loss of molecules. \textbf{b, c,} The molecule number (\textbf{b}) and temperature (\textbf{c}) are measured at various hold times after a 100-ms-long exponential evaporation ramp to a trap power of $0.21\; {\rm W}$ ($7\;{\rm \mu K}$ trap depth), followed by a 10-ms-long recompression to $1.2\;{\rm W}$ ($40\;{\rm \mu K}$ trap depth; trap-depth values are for NaLi throughout). In the number plot (\textbf{b}) the red dashed line is an exponential loss fit and the blue solid line is a two-body loss fit (details in the main text). The dashed line in the temperature plot (\textbf{c}) is an exponential fit. No temperature drop occurs in the absence of Na atoms. The exponential fit of the molecule temperature agrees well with the measurement but should be regarded as an interpolation to determine the initial slope of the thermalizaion. For this measurement, the Na number was about $1.5\times 10^{5}$. Data values represent the average and error bars represent the standard error of the mean, estimated from the fitting and statistical errors of 3–8 measurements.}
		\label{fig:thermalization}
	\end{figure}
    Our experimental setup is summarized in Fig. 1. Similarly to our previous work\cite{NaLiFesh, NaLiGround}, we produce about $3.0\times10^{4}$ NaLi molecules in the rovibrational ground state of the triplet potential using a mixture of Na and Li atoms (further details in Methods). We also prepare about $1.0\times10^5$ Na atoms in the upper stretched hyperfine state (in the low-field basis, $\ket{F,m_{F}}=\ket{2,2}$, where $F$ and $m_{F}$ are the hyperfine and magnetic quantum numbers, respectively) in a one-dimensional (1D) optical lattice formed by 1,596-nm light. Owing to the differential polarizability at 1,596-nm, molecules feel a deeper trapping potential than atoms. This results in a sudden increase of the potential energy as atoms associate and form molecules (see Fig. 2a). Immediately after production, the effective temperature of the molecules is $2.80(6)\;\mu {\rm K}$ and the temperature of Na atoms is $2.42(3)\;\mu {\rm K}$ (all uncertainties are as defined in the legend of Fig. 2). As the molecules thermalize with the Na atoms and a hot fraction of atoms evaporates out of the trap, the temperatures of both particles settle to $2.23(6)\;\mu {\rm K}$ (see Methods for molecular thermometry).
    
	Although this initial settling of temperatures hints at sympathetic cooling, we are able to see much stronger effects by forced cooling and heating of Na atoms. We evaporate Na atoms with almost no loss of molecules by taking advantage of the particles’ different polarizabilities: $\alpha_{\text{NaLi}}/\alpha_{\text{Na}} = (m\omega^{2})_{\text{NaLi}}/(m\omega^{2})_{\text{Na}} \approx 2.6$, where $\alpha_i$ is the polarizability of particle $i$, $\omega$ is the angular frequency for oscillation in the trap and $m$ is the particle mass. The curve in Fig. 2c shows the thermalization between molecules and atoms with a large initial temperature difference after an exponential evaporative cooling ramp followed by a recompression of the trap (see Fig. 2 legend). As we hold the particles in the trap, their temperatures approach each other. Owing to the large particle number ratio, $N_{\text{\rm Na}}/N_{\text{\rm NaLi}}\approx 7$, the molecule temperature decreases by $0.68(9) \; \mu{\rm K}$ after thermalization, whereas the temperature of Na atoms increases only by $70(30) \; {\rm nK}$. However, if the Na is removed immediately before the hold time, the molecule temperature remains fixed during the same period.
	
	As further evidence of thermalization, the sympathetic heating of molecules with hot atoms is shown in Fig. 3. We first prepare the atom–molecule mixture fully thermalized at $1.5 \; \mu{\rm K}$, after a 10-ms-long exponential evaporative ramp, followed by a 200-ms-long recompression of the trap to the initial trap power of $1.5\; {\rm W}$ ($50 \; \mu{\rm K}$ trap depth). Then, we selectively heat the atoms to $2.07(5) \; \mu{\rm K}$ by sinusoidally modulating the trap amplitude at twice the sodium trap frequency ($2 \omega_{Na}= 2\pi\times920\; {\rm Hz}$) with a modulation amplitude of $20 \%$ of the trap depth for 100 ms. After 200 ms, particles thermalize and the molecule temperature rises to the temperature of the heated Na atoms. When the trap amplitude is modulated in the same manner without the Na atoms, the temperature of molecules remains at $1.59(5) \; \mu{\rm K}$. The heating process for sodium also induces centre-of-mass motion and breathing oscillations that cause the Na density to depend on time in the early stages of sympathetic heating, which is the reason for the delay in heating and for the non-exponential thermalization curve.
	\begin{figure}[H]
	    \centering
		\includegraphics[width = 98mm, keepaspectratio]{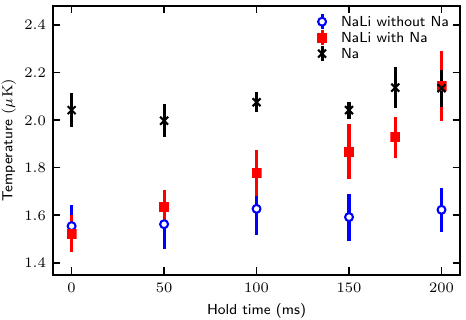}
		\caption[width = \textwidth]{\textbf{Sympathetic heating.} After forced heating of Na atoms (see main text for details), the NaLi molecule temperature (red squares) rises and reaches that of Na atoms (black asterisk) as both particles thermalize. The temperature of molecules without hot atoms (blue circles) remains low. Values and error bars are estimated as in Fig. 2 from four measurements.}
		\label{fig:heating}
	\end{figure}
	
	\begin{figure}[H]
		\centering
		\includegraphics[width = 98mm, keepaspectratio]{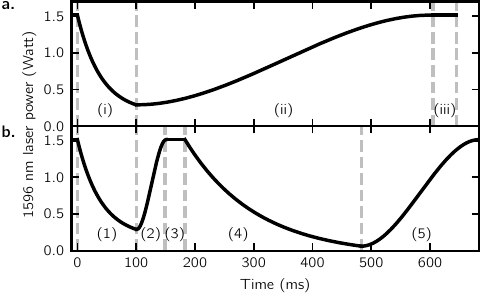}
		\caption{\textbf{Evaporation sequences.} The initial power of the 1,596-nm trapping laser is $1.5\; {\rm W}$. At the end of evaporation we recompress the trap to the initial power to increase the thermalization rate and for a straightfoward comparison of the NaLi density without re-scaling the trap volume. \textbf{a}, Single evaporation: (i) exponential forced evaporation ($\tau =$ 40 ms) for 100 ms with trap depth $U(t) = A\exp (-t/\tau)+B$, where $A$ and $B$ are determined from the trap depths at $t =$ 0 and 100 ms; (ii) recompression for 500 ms; (iii) hold for 40 ms for complete thermalization. \textbf{b}, Double evaporation: (1) exponential forced evaporation ($\tau =$ 40 ms) for 100 ms; (2) recompression to the initial trap depth for 50 ms; (3) hold for 30 ms; (4) exponential evaporation ($\tau =$ 120 ms) for 300 ms; (5) recompression for 200 ms.}
		\label{fig:evap_seq}
	\end{figure}
	To measure the rate of thermalization, we return to the simpler situation of cooling in Fig. 2. We fit the temperature to a simple exponential model, $T(t) = (T_{0}-T_{\infty})\exp(-\Gamma_{\rm th} t) + T_{\infty}$, where $T_{0}$ and $T_{\infty}$ represent the initial and infinite-time-limit temperature of molecules, respectively, and obtain the thermalization rate, $\Gamma_{\text{th}}=26(5) \; \text{s}^{-1}$. Considering that thermalization requires about $3/\xi$  collisions (for Na–NaLi, $\xi \approx 1$; see Methods), we obtain the average elastic collision rate per particle, $\Gamma_{\text{el}}$, from the measured thermalization rate: $\Gamma_{\text{el}} \approx (3/\xi)\Gamma_{\text{th}} = 80(14)\; \text{s}^{-1}$. In the presence of Na atoms, the initial loss rate of molecules is $\Gamma_{\text{inel}} = 1.29(6) \; \text{s}^{-1}$, as obtained from a fit to the exponential loss model $N(t)=N_0 \exp(-\Gamma_{\text{inel}} t)$ (red dashed line in Fig. 2b). Comparing the average elastic collision rate to the total loss rate, we obtain the ratio of elastic to inelastic collisions for NaLi, $\gamma\gtrsim \Gamma_{\text{el}}/\Gamma_{\text{inel}} = 62(12)$. Without Na atoms, the molecular loss follows a two-body loss model, $N(t) = N_0/(\beta t + 1)$ (blue solid line in Fig. 2b), from which we obtain an initial loss rate of $\beta = 1.02(6) \; \text{s}^{-1}$. By considering the difference between $\Gamma_{\text{inel}}$ and $\beta$ as the effective inelastic loss rate for collisions between Na and NaLi, we obtain the ratio of ``good'' to ``bad'' collisions, $\gamma \approx \Gamma_{\rm el}/(\Gamma_{\rm inel} - \beta)\approx300$ with an uncertainty of $40 \%$. The Na–NaLi loss rate constant is $4.0 \pm 1.3 \times 10^{-13}\; {\rm cm^{3}\ s^{-1}}$; this is more than two orders of magnitude smaller than the universal loss model rate constant for Na–NaLi $s$-wave collisions\cite{univlossrate}, which is $1.7 \times 10^{-10}\; {\rm cm^{3}\ s^{-1}}$ (see Methods). Whereas Na and NaLi in their upper stretched states form a stable mixture, as shown, preparing Na atoms in their lowest hyperfine state ($\ket{F,m_{F}}=\ket{1,1}$) gives a loss rate consistent with the universal loss model\cite{NaLiGround}.

    We consider the two different evaporation ramps shown in Fig. 4. The molecule numbers, temperatures and PSDs resulting from these ramps are displayed in Fig. 5. For a single-species system with losses, optimal evaporation can be achieved by a single, continuous decrease of the trap depth. We achieve an increase in PSD by a factor of $7$ by using a single stage of evaporation (Fig. 4a, red squares in Fig. 5). This increase is limited by the low initial Na density. In our system, thermalization is dominated by the Na density, whereas loss is dominated by the NaLi density. To overcome the low Na density, we shorten the initial evaporation and recompression cycle ((1) and (2) in Fig. 4b). This cools the Na atoms and quickly increases the density of the cold Na for fast thermalization. After the cold, dense Na efficiently pre-cools the molecules during a short hold of 30 ms in the tight trap (3), we apply the second evaporation ramp (4). In this double evaporation, we achieve a peak PSD of $2.0(2)\times10^{-2}$ (blue circles in Fig. 5), which is $20$ times higher than the initial PSD. Before the final recompression of the trap, the lowest temperature of the molecules is $220(20)\;{\rm nK}$.
	\begin{figure}[H]
		\centering
		\includegraphics[width = 98mm, keepaspectratio]{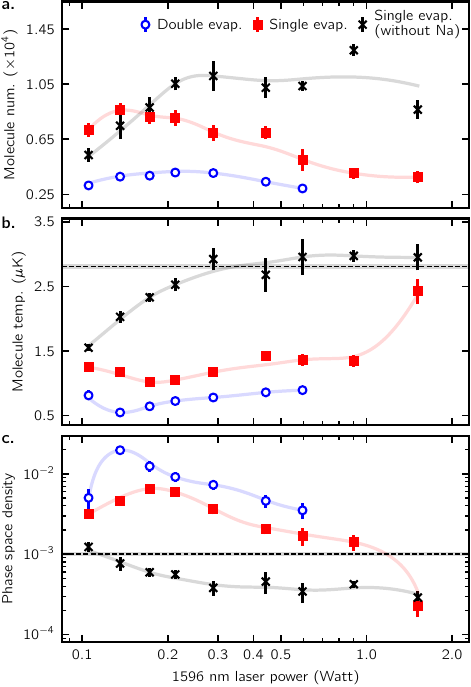}
		\caption{\textbf{Increasing phase-space density.} \textbf{a–c}, The number (\textbf{a}), temperature (\textbf{b}) and PSD (\textbf{c}) of NaLi molecules are plotted versus the power of the 1,596-nm trapping laser at the end of the forced exponential evaporation ((i) for single evaporation and (1) for double evaporation in Fig. 4). For the double evaporation, we ramp the power down to $0.06\; {\rm W}$ ($2\;{\rm \mu K}$ trap depth) in (4), the lowest point that gives a molecular signal high enough for consistent temperature measurement. Single-stage evaporation to this low trap depth leads to loss of almost every molecule. The black dashed lines in \textbf{b} and \textbf{c} indicate the molecule temperature and PSD after STIRAP and removal of free atoms (shaded areas show $\pm$ 1 error bar). This PSD corresponds to $2.6(1) \times 10^4$ molecules. When Na atoms are not in the trap or when the evaporation is not performed to a sufficiently low trap depth, the PSD is lower than the initial value. This is mainly due to the molecular two-body loss. The solid curves are guides for the eye. Values and error bars are estimated as in Fig. 2 from 4–9 measurements.}
		\label{fig:psd_evap}
	\end{figure}
	With Na present, more molecules survive the evaporation sequence as we evaporate Na atoms to a lower trap depth. This is mainly due to suppressed two-molecule loss. The loss rate constant scales linearly with the temperature owing to the p-wave character of the collisions\cite{univlossrate}. Evaporation to too low trap depth decreases the Na density, which makes the sympathetic cooling of molecules inefficient, and anti-evaporation (heating due to the preferential loss of the coldest molecules) dominates. This appears as a temperature increase below a trap power of $0.17\; {\rm W}$ ($5.7\;{\rm \mu K}$ trap depth) for a single stage of evaporation, or $0.14\; {\rm W}$ ($4.5\;{\rm \mu K}$ trap depth) for two-stage evaporation. Without Na, the ramp-down in trap depth, (i), is adiabatic to a trap power of around $0.3\; {\rm W}$, as evidenced by the constant molecule numbers and temperatures. Below that trap power, the ramp-down starts becoming non-adiabatic. As a result, a hot fraction of molecules escape, and both the molecule number and the temperature start decreasing. We note that because fermionic molecules do not thermalize by themselves, this temperature should be regarded as the average kinetic energy of molecules that are not in thermal equilibrium.
	
	The favourable collisional properties of the spin-polarized Na–NaLi mixture in fully stretched hyperfine states result from strong suppression of electronic spin flips during collisions, which could otherwise lead to fast reactive or inelastic losses. In fully stretched states, direct spin exchange is forbidden and residual spin flips can occur only by weaker interactions. The main contribution to the weak spin flips is from dipolar relaxation due to direct coupling between the magnetic dipole moments of the spins of Na and NaLi. Studies in other spin-polarized systems\cite{DoyleNHpN, TscherbulSrFSymp, TscherbulCaHLi} predict spin flips induced by anisotropic electrostatic interaction at short range and intramolecular spin–spin and spin–rotation couplings to be weaker at ultracold temperatures. Dipolar relaxation can be eliminated by using the strong-field-seeking stretched state, but this was not necessary in our work. It is possible that the Na–NaLi system is favourable owing to the low reduced and molecular masses and the small relativistic effects, which result in a large rotational constant\cite{WesleySpinRelax, collisionTheoryKrems}, low density of states\cite{StickyBohn} and small spin–orbit coupling\cite{SpinOrbitSpinRelax}, respectively. Further, ref. \cite{NHChemReacts} has shown that intramolecular spin–spin coupling causes spin flips, leading to chemical reactions for triplet NH molecules, thwarting evaporative cooling. Our results show a clear need for further work to determine what other molecules---and what conditions of internal states and external fields---would be suitable for collisional cooling at ultracold temperatures.
	
	Technical upgrades to our apparatus should allow cooling into the quantum degenerate regime by increasing the initial Na density. The dominant loss mechanism---$p$-wave collisions between molecules---is expected to slow down compared to thermalization at lower temperatures, improving the efficiency of cooling. Further, the atomic and molecular states used in this work are magnetically trappable. Magnetic trapping would allow the use of much larger samples with perfect spin polarization. This would also eliminate the major concern of molecular loss induced by trapping light\cite{KarmanMolLoss}, making NaLi an ideal system for studying quantum chemistry. Concerns about sticky collisions\cite{StickyBohn}, which can lead to a long-lived collision complex with high density of states, have led to pessimism over the prospects of achieving collisional cooling of ultracold molecules. This work demonstrates that sticky collisions do not limit NaLi, and probably other low-mass systems, suggesting a bright future for cooling light molecules.

    \newpage
    % \begin{nolinenumbers}
    % \textbf{References}

    \newpage
    \begin{center}
        \textbf{Methods} \\
    \end{center}
	\textbf{Atomic sample preparation} 
	We produce an ultracold mixture of $^{23}{\rm Na}$ and $^6{\rm Li}$ in their upper stretched hyperfine states (in the $\ket{F,m_{F}}$ basis, these states correspond to $\ket{2,2}$ for Na and $\ket{3/2,3/2}$ for Li) in an Ioffe–Pritchard magnetic trap at a bias field of $2\;{\rm G}$. We cool the Na atoms using microwave-induced evaporation and sympathetically cool the Li atoms with the Na atoms\cite{SympatheticLiCool,WiemanSympBECs, LiCsmixture, GuptaSymp}. Then, the mixture is transferred into the combination of two coaxial optical traps: a 1,064-nm optical dipole trap ($30\;\mu{\rm m}$ waist, $10\;{\rm W}$ power) and a 1,596-nm 1D optical lattice ($50\;\mu{\rm m}$ waist, $2\;{\rm W}$ power). After 0.6 s of forced evaporation in the 1,064-nm optical dipole trap, the trap is switched off. Then, in the 1,596-nm 1D lattice, the sample is transferred to the lowest-energy Zeeman states ($\ket{1,1}$ for Na, and $\ket{1/2,1/2}$ for Li) using a Landau–Zener magnetic field sweep. By controlling the microwave power for the Landau–Zener sweep, we intentionally leave a fraction of Na atoms in the upper stretched hyperfine state for later use as coolants of the NaLi molecules.
	\\ \\
	\textbf{Molecule formation and detection}
	As previously demonstrated with other atomic species\cite{JinYeKRb, DenschlagRb_2, NagerlCs_2, NagerlRbCs, CornishRbCs, ZwierleinNaK, WangNaRb, NaLiGround, ZhaoNaKFesh, BlochNaK}, NaLi ground-state molecules are created from Feshbach molecules in the least-bound vibrational state. In a 1,596-nm 1D optical lattice, NaLi Feshbach molecules form in the $a^{3}\Sigma^{+}$, $v=10$, $N=2$, $m_{N}=-2$ state ($v$ and $N$ are the vibrational and rotational molecular quantum numbers, respectively, and $m_{N}$ is the magnetic quantum number of $N$) from the atoms in the lowest-energy Zeeman states by sweeping the magnetic field across an interspecies scattering resonance (Feshbach resonance) at 745 G (ref. \cite{NaLiFesh}). Then, about $98 \%$ of Feshbach molecules are transferred to the triplet ground state ($a^{3}\Sigma^{+}$, $v=0$, $N=0$) by means of a 30-$\mu{\rm s}$-long STIRAP sequence with an intermediate state ($v=11$, $N=1$, $m_{N}=-1$) in the $c^{3}\Sigma^{+}$ excited potential. The use of 1,596-nm light for the trap improves the lifetime of the NaLi molecules by suppressing spontaneous photon scattering compared to the more common 1,064-nm trap. The 1D lattice provides strong axial confinement to overcome the anti-trapping produced by magnetic curvature\cite{NaLiGround}. The STIRAP light (upleg: $833\;{\rm nm}$, $360.00381(1)\;{\rm THz}$; downleg: $819\;{\rm nm}$, $366.23611(1)\;{\rm THz}$)\cite{NaLiGround,NaLiPA,NaLiTwoPhoton} is obtained from two home-built external cavity diode lasers (ECDL)\cite{ECDL} locked to an ultralow-expansion cavity. The relative linewidth of the two ECDLs is less than $1\;{\rm kHz}$. To attain sufficient optical power for the upleg transition, we use a 500-mW high-power diode laser injection-locked by 2 mW of ECDL light. After molecule formation, a resonant light pulse is applied for about 1 ms to remove the free atoms that were not converted to molecules. These atoms in the lowest-energy Zeeman states accelerate the loss of NaLi to roughly the universal loss model rate\cite{NaLiGround}. Sodium atoms in this state are optically pumped by the resonant removal light into a dark Zeeman state ($\ket{F,m_{F}} = \ket{2,1}$). Adding a repump beam to address this state removes the need to reduce the magnetic field (by instantaneously shutting off the current in the coils generating the 745-G field), which caused considerable heating and loss of molecules in our previous work\cite{NaLiGround}. We detect ground-state molecules using reverse STIRAP and a magnetic field sweep across the Feshbach resonance followed by resonant absorption imaging of the resulting free atoms.
    \\ \\
	\textbf{Thermometry of molecules} To avoid issues with different atomic and molecular trapping potentials, we dissociate molecules in free space after turning off the trap, which allows accurate molecular thermometry. Given that the particle velocities are fixed after the trap is turned off, the density at some time of flight maps the velocity distribution of molecules in the trap. The molecule temperature is determined from a fit to such density profiles. Extra kinetic energy could be added in quadrature during the dissociation of Feshbach molecules into atomic continuum states\cite{dissociationKE}. We determine the parameters of Feshbach dissociation (that is, the rate of the magnetic field sweep) to minimize the released dissociation energy. All uncertainties quoted in this article are from statistical and fitting errors. The systematic uncertainty, which is mainly determined from the imaging magnification, is about $6 \%$ but has no effect on the measured thermalization rates because it gives a common multiplier to all temperature measurements.
   \\ \\
   \textbf{Na-NaLi $s$-wave elastic scattering cross-section} The elastic scattering rate between Na and NaLi is given by $\Gamma_{\rm el} = n_{\rm ov}\sigma_{\rm el}v_{\rm rel}$ where $n_{\rm ov}$ is the overlap-averaged density of the mixture, $\sigma_{\rm el}$ is the elastic-scattering cross-section, and $v_{\rm rel}$ is the mean relative velocity\cite{mixtureCrosssec}:
	\begin{equation}
	\begin{aligned}
	    v_{\rm rel} &= \sqrt{\frac{8k_B}{\pi} \bigg(\frac{T_{\text{Na}}}{m_{\text{Na}}} + \frac{T_{\text{NaLi}}}{m_{\text{NaLi}}}  \bigg)} \\
	    n_{\rm ov} &= \frac{I}{\mfrac{N_{\text{Na}} N_{\text{NaLi}}}{N_{\text{Na}}+N_{\text{NaLi}}}} 
	    = \frac{N_{\text{Na}}  N_{\text{NaLi}} \bar{\omega}_{\text{Na}}^3\! \left[\!\left(\frac{2\pi k_B T_{\text{Na}}}{m_{\text{Na}}}\right)\!\!\! \left(1 + \frac{\alpha_{\text{Na}}}{\alpha_{\text{NaLi}}}\frac{T_{\text{NaLi}}}{T_{\text{Na}}}\right) \!\right]^{-\frac{3}{2}}}{\mfrac{N_{\text{Na}} N_{\text{NaLi}}}{N_{\text{Na}}+N_{\text{NaLi}}}}
	\end{aligned}
	\end{equation}
	where $m_{i}$, $T_{i}$, and $\alpha_{i}$ represent the mass, temperature, and polarizability of the particle ${i}$ respectively, $I=\int dV n_\text{Na} n_\text{NaLi}$ is the density overlap integral of the mixture in a harmonic trap, and $\bar{\omega}_{\text{Na}} = (\omega_x\omega_y\omega_z)^{1/3}=2\pi \times (540 \cdot 410 \cdot 34000)^{1/3}\;  \text{Hz}$ is the geometric mean of the trap frequencies, and $k_B$ is Boltzmann constant. The differential gravitational sag between molecules and atoms has a negligible effect on the overlap density. The effective number of particles per lattice site, $N_{(\text{Na, NaLi})}$, is the total number of particles divided by a factor $l_{\text{eff}}/a$, where $a = \lambda/2$ is the lattice spacing ($\lambda$, wavelength of the trap laser) and $l_{\text{eff}}$ is the effective axial width (in the $z$-direction, see Fig. 1) of the atomic density distribution across all lattice sites:
	\begin{equation}
	    l_{\text{eff}} = 2 \; \frac{ \int_{0}^{\infty} z \exp(-z^2/2\sigma^2)dz}{ \int_{0}^{\infty} \exp(-z^2/2\sigma^2)dz} =\sqrt{\frac 8 \pi}\sigma ,
	\end{equation}
	where $\sigma=0.72(1)\; {\rm mm}$ is the Gaussian width, obtained from a fit of the atomic density profile.	
    
    In a mass-imbalanced system, the factor $3/\xi$ quantifies the approximate average number of collisions per particle required for thermalization, where $\xi =  4m_{\text{Na}}m_{\text{NaLi}}/(m_{\text{Na}}+m_{\text{NaLi}})^2$ (refs. \cite{LiCsmixture}). For the Na–NaLi mixture, this corresponds to approximately three collisions because the mass difference is small. Thus, the relation between the thermalization rate and the elastic-scattering rate is given by $\Gamma_{\text{th}} \approx  \Gamma_{\text{el}}/(3/\xi)$. In our system, where the particle number is largely imbalanced (that is, $N_{\text{\rm Na}}/N_{\text{\rm NaLi}}\approx 7$), we can write the thermalization rate as:
    
    \begin{equation}
	    \Gamma_{\rm th} \approx \frac{(I/N_{\text{NaLi}})\sigma_{\rm el}v_{\rm rel}}{3/\xi},    
	\end{equation}
	where $I/N_{\text{NaLi}}$ is the average density of Na atoms seen by NaLi molecules. Given the measured thermalization rate, we obtain the elastic-scattering cross-section between a molecule and an atom, $\sigma_{\rm el} \approx (3/\xi)(N_{\text{NaLi}}/v_{\rm rel}I)\Gamma_{\rm th}= 2.4(5) \times 10^{-11} {\rm cm}^{2}$ and the corresponding scattering length, $a = 263(24)a_0$ where $a_0$ is the Bohr radius. 

	In addition to the simple exponential fit, we perform a numerical simulation of the thermalization process based on the differential equation for $\Delta T$ = $T_{\rm NaLi}$ - $T_{\rm Na}$ (refs. \cite{GuptaSymp, mixtureCrosssec, LiCsmixture}):
	
	\begin{equation}\label{eq:diff}
       \frac{1}{\Delta T}\frac{d(\Delta T)}{dt} = -\Gamma_{\rm th} = -\frac{\xi}{3}n_{\rm ov}\sigma_{\rm el}v_{\rm rel}.
   \end{equation}
   
   In the numerical simulation, the temperature dependence of $n_{\rm ov}$, $\sigma_{\rm el}$, and $v_{\rm rel}$ is fully taken into account. The s-wave elastic-scattering cross-section between Na and NaLi is given by $\sigma_{\rm el} = \mfrac{4\pi a^2}{[1 -(1/2)k^2r_{0}a]^2 + (ka)^2}$, where $\hbar k$ = $\mu v_{\rm rel}$ is the relative momentum ($\mu$ is the reduced mass, $\hbar$ is the reduced Planck constant) and $r_{0}$ is the effective range calculated from the scattering length, $a$, and from $C_6^{\text{Na-NaLi}}$ (see Methods section ``Na-NaLi loss rate constant from the universal loss model''). We iteratively solve equation (4) by varying $a$, and find the optimal solution that minimizes the weighted sum of squared residuals: $\Sigma_{i} [(\Delta T_{m,i} - \Delta T_{n,i})/\sigma_{\Delta T_{m,i}}]^2$, where the subscript $n$ denotes the numerical solution, $m$ denotes the measurement and $\sigma_{\Delta T_{m,i}}$ is the uncertainty of the measurement at the $i^{\rm th}$ time step, estimated as in Fig. 2. The numerical solution for $a$ obtained in this way agrees with the result calculated from the simple exponential model described above within the uncertainty.
    \\ \\
    \textbf{Na-NaLi loss rate constant} The Na-NaLi loss rate constant is defined as the differential loss rate, $\Gamma_{\rm inel} - \beta$, normalized by the density of Na overlapped with NaLi (that is, the overlap-averaged density of the mixture, $n_{\rm ov}$), where $\Gamma_{\rm inel}$ is the molecular loss rate with Na and $\beta$ is without Na. As shown in the calculation of the $s$-wave elastic-scattering cross-section between Na and NaLi, $n_{\rm ov} \approx I/N_{\rm NaLi}$, given the large imbalance of particle numbers. Thus, the Na-NaLi loss rate constant mentioned in the main text is calculated from $(\Gamma_{\rm inel} - \beta)/(I/N_{\rm NaLi})$.
	\\ \\ 
	\textbf{Na-NaLi loss rate constant from the universal loss model} We compare the measured loss rate constant for Na and NaLi with the theoretical prediction based on the universal loss model\cite{univlossrate}. When calculating the theoretical loss rate constant, we estimate the van der Waals coefficient between Na and NaLi, $C_{6}^{\text{Na-NaLi}}$, as the sum of the $C_6$ coefficients for the atomic pairs that constitute the collision partners, using values from ref. \cite{dispersioncoeff}: $C_{6}^{\text{Na-NaLi}}$ = $C_{6}^{\text{Na-Li}}$ + $C_{6}^{\text{Na-Na}}$.
    
    \bigbreak
    
    \newcounter{methodCounter}
    \setcounter{methodCounter}{41}

    \newpage
    \noindent
    \textbf{Acknowledgements} \\
    We thank M. Zwierlein for discussions and J. Yao for technical assistance. We acknowledge support from the NSF through the Center for Ultracold Atoms and award 1506369, from the NASA Fundamental Physics Program and from the Samsung Scholarship.
    \\ \\
    \textbf{Contributions} \\
    H.S., W.K. and A.O.J. conceived the experiment. H.S. led the data measurement. H.S. and A.O.J. performed the data analysis. H.S., J.J.P. and A.O.J. designed and constructed the experimental setup. All authors discussed the results and contributed to the writing of the manuscript.
    \\ \\
    \textbf{Data availability} \\
    The data that support the findings of this study are available from the corresponding author upon reasonable request.
    \\ \\
    \textbf{Competing interests} \\
    The authors declare no competing financial interests.
    \\ \\
    \textbf{Corresponding author} \\
    Correspondence to Hyungmok Son (e-mail: hson@g.harvard.edu)

\end{document}